\documentclass[%
 aip,
 amsmath,amssymb,
 reprint,%
]{revtex4-1}

\usepackage{graphicx}
\usepackage{dcolumn}
\usepackage{bm}

\usepackage[utf8]{inputenc}
\usepackage[T1]{fontenc}
\usepackage{mathptmx}
\usepackage{etoolbox}
\usepackage{amsmath}
\usepackage{amssymb}
\def \f {\text{f}}
\def \g {\text{g}}

\def \p {\text{p}}

\def \s {\text{s}}
\def \b {\text{b}}

\def \i {\text{i}}
\def \e {\text{e}}
\def \q {\text{q}}
\def \sl {\text{sl}}
\def \fe {\text{fe}}
\def \sh {\text{sh}}

\def \as {\text{as}}

\def \I {\text{I}}
\def \II {\text{II}}
\def \III {\text{III}}

\def \b {\text{b}}
\def \th {\text{th}}
\def \pl {\text{pl}}
\def \be {\begin{equation}}
\def \ee {\end{equation}}
\draft 

\begin{document}


\title{Self-similarity of the third type in ultra relativistic blastwaves} 

\author{Tamar Faran}\email{tamar.faran@princeton.edu}\affiliation{Department of Astrophysical Sciences, Princeton University, Princeton, NJ 08544, USA}
\author{Andrei Gruzinov}\affiliation{CCPP, Physics Department, New York University, 726 Broadway, New York, NY 10003}
\author{Re'em Sari}\affiliation{Racah Institute of Physics, Hebrew University, Jerusalem 91904, Israel}






\begin{abstract}
A new type of self-similarity is found in the problem of a plane-parallel, ultra-relativistic blast wave, propagating in a powerlaw density profile of the form $\rho \propto z^{-k}$. Self-similar solutions of the first kind can be found for $k<7/4$ using dimensional considerations. For steeper density gradients with $k>2$, second type solutions are obtained by eliminating a singularity from the equations. However, for intermediate powerlaw indices $7/4<k<2$ the flow does not obey any of the known types of self-similarity. Instead, the solutions belong to a new class in which the self-similar dynamics are dictated by the non self-similar part of the flow. We obtain an exact solution to the ultra-relativistic fluid equations and find that the non self-similar flow is described by relativistic expansion into vacuum, composed of (1) an accelerating piston that contains most of the energy and (2) a leading edge of fast material that coincides with the interiors of the blastwave and terminates at the shock. 
The dynamics of the piston itself are self-similar and universal, and do not depend on the external medium. The exact solution of the non self-similar flow is used to solve for the shock in the new class of solutions.
\end{abstract}


\maketitle 

\section{Introduction}
Self-similar solutions offer great mathematical simplicity by reducing a set of partial differential equations (PDEs) to a set of ordinary differential equations (ODEs). They are used to characterize the asymptotic behaviour of a physical system when its initial scales become unimportant, and can sometimes be described by analytical solutions. Self-similar solutions are traditionally classified into one of two types: when the self-similar properties of the system can be found from dimensional considerations, the solutions are of first type. Second type solutions are obtained when one is required to solve an eigenvalue problem (\citealt{Zeldovich1967}, \citealt{Barenblatt1976}).

A canonical example for self-similarity in hydrodynamics is the strong explosion problem. Consider a powerlaw density profile of the form $\rho \propto r^{-k}$, where $r$ is the distance from the origin and $k$ is a constant. The release of a large amount of energy at $r=0$ produces a strong shock wave that scales with time like $R \propto t^\alpha$, where $\alpha$ is a function of $k$. Dimensional analysis can be used when the energy is a relevant parameter of the self-similar flow, such that $\alpha$ is found by enforcing energy conservation. First type solutions for Newtonian spherically symmetric explosions were originally obtained by \citealt{Taylor}, \citealt{Neumann} and \citealt{Sedov}. \citealt{Waxman1993} showed that when the density falls sufficiently fast ($k>3$) dimensional considerations give a wrong temporal scaling for the shock as it accelerates and loses causal contact with the bulk of the flow. Second type solutions are then obtained for $k>3.26$ by eliminating a singularity from the equations, corresponding to a smooth crossing of the sonic line. \citealt{Waxman1993} concluded that when $3<k<3.26$ the solutions cannot be described by either of the known types of self-similarity. \citealt{Gruzinov2003} subsequently showed that the solutions in that parameter space belong to a new class (see also \citealt{Kushnir2010}). The uniqueness of the new class will be explained when discussing its ultra relativistic analogue.


\citealt{BM76} obtained self-similar solutions of the first type for ultra-relativistic blastwaves, produced when the explosion energy is much greater than the rest mass of the exploding material. The blastwave is engulfed by a shock wave with a Lorentz factor $\Gamma \gg 1$, assumed to follow a temporal scaling characterized by the self-similar index 
\begin{equation}
    m = -\frac{d \log \Gamma^2}{d \log t}\,.
\end{equation}
The pioneering work of \citealt{BM76} was followed by many other studies of first and second type solutions in various geometries (e.g., \citealt{Best2000,Perna2002,Hidalgo2005,Sari2006}).

This work is focused on ultra relativistic blastwaves in plane parallel geometry. We consider an ultra relativistic shock travelling in the powerlaw density profile given by
\begin{equation}\label{eq:RhoPowerLaw}
    \rho \propto z^{-k}\,,
\end{equation}
where $z$ is the planar spatial coordinate. By studying the solutions at large distances behind the shock we find that the pressure and Lorentz factor asymptotically approach power-laws in time and position. Three different types of self-similar solutions are then inferred from the asymptotic flow. When the energy in the asymptotic solution converges, the solution is of first type ($k<7/4$). If the asymptotic solution, when applied to the entire space, has diverging energy, the dynamics cannot be inferred from energy conservation considerations. We will show that two different classes of solutions can describe such flows; the first is the well known second type solutions, in which the self-similar part of the flow is not in causal contact with its non self-similar part due to the existence of a sonic point ($k>2$). The similarity exponent is found by requiring that the solution smoothly crosses the sonic line. In the absence of a singularity, the solution is neither of first nor of second type, defining a solution gap ($7/4<k<2$). In this region, we find a new class of self-similar solutions, in which the dynamics of the shock are dictated by the rear region of the flow that is not part of the same self-similar solution. A solution for the non self-similar flow, which behaves as expansion into vacuum, is required in order to solve for the similarity index $m$. 
As already mentioned, the new class of solutions was first identified in Newtonian blastwaves by \citealt{Gruzinov2003}, and was termed `self-similarity of the third type'. However, the Newtonian case is a trivial manifestation of third type solutions; the shock is driven by a piston of infinite mass and momentum, and simple considerations show that both the piston and the shock must travel at a constant velocity. In this work, we introduce a non-trivial example of the new type of self-similarity in the ultra-relativistic regime.

The paper is organized as follows: after deriving the self-similar equations in \S \ref{sec:SelfSimEq}, we give an overview of first and second type solutions in the ultra-relativistic blastwave problem in \S \ref{sec:FirstSecondTypeSol}.
We then study the asymptotic behaviour of the flow at large distances behind the shock in \S \ref{sec:AsymptoticPLSolution} and obtain exact solutions for expansion into vacuum in \S \ref{sec:ExpIntoVacuum}. The self-similar scaling of the shock in new class of solutions is found in \S \ref{sec:Shock}. We compare to numerical simulations in \S \ref{sec:Simulations} and summarize in \S \ref{sec:Summary}.

\section{The self similar equations}\label{sec:SelfSimEq}
This section broadly follows the formalism of \citealt{BM76} in plane-parallel geometry. We consider a fluid with an ultra-relativistic equation of state, such that the internal energy and pressure satisfy the relation $p = e/3$. As a consequence, the velocity and pressure of the flow are decoupled from the material density, simplifying the problem to the conservation of energy and momentum alone:
\begin{subequations} \label{eq:hydro_ultra_rel1}
    \begin{equation}
       \frac{\partial}{\partial t} \gamma^2(e+\beta^2p)+\frac{\partial}{\partial z}  \gamma^2 \beta (e+p) = 0
       \end{equation}
       \begin{equation}
       \frac{\partial}{\partial t}  \gamma^2 \beta (e+p) + \frac{\partial}{\partial z} \gamma^2 \beta^2 (e+p) + \frac{\partial p}{\partial z}=0\,,
       \end{equation}
\end{subequations}
where $\gamma$ is the Lorentz factor of the flow and $\beta = \sqrt{1-\gamma^{-2}}$ is its associated reduced velocity.
Equation set \eqref{eq:hydro_ultra_rel1} can be rewritten in a more compact form by making the transformation to a new variable,
\begin{equation}
    x \equiv t-z\,.
\end{equation}
Keeping only leading orders of $1/\gamma^2$ and defining 
\begin{equation}
    q \equiv \gamma^2\,,
\end{equation}
the conservation equations become
\begin{subequations} \label{eq:hydro_ultra_rel}
    \begin{equation}
       4\frac{\partial q p}{\partial t} + \frac{\partial p}{\partial x} = 0 
       \end{equation}
       \begin{equation}
       \frac{\partial p}{\partial t} + \frac{\partial p/q}{\partial x} = 0\,.
       \end{equation}
\end{subequations}
Consider a flow with a characteristic Lorentz factor $\Gamma$ and an associated characteristic position $R$, related by $\Dot{R} = \sqrt{1-1/\Gamma^2} \simeq 1-1/(2\Gamma^2)$, where the second equality assumes the ultra-relativistic limit.
In the context of ultra-relativistic blastwaves, $R$ and $\Gamma$ are taken to be the position and Lorentz factor of a shock wave, respectively, and the typical scale height of the flow in the immediate downstream of the shock is of order $R/\Gamma^2$. It is then natural to define the following similarity variable
\begin{equation} \label{eq:chi_def}
    \chi = 1+2(m+1)\frac{x}{R/\Gamma^2} \,,
\end{equation}
which places the shock at $\chi = 1$. The shock is assumed to propagate in a density profile defined by Eq \eqref{eq:RhoPowerLaw}. The profiles of $q$ and $p$ behind the shock are given by the self-similar functions $g(\chi)$ and $f(\chi)$, defined by
\begin{subequations}\label{eq:selfsim_func_def}
\begin{equation} \label{eq:gamma_def}
    q(x,t) = \frac{\Gamma^2(t)}{2}g(\chi)
\end{equation}
\begin{equation}\label{eq:p_def_UR}
    p(x,t) = P(t)f(\chi)\,.
\end{equation}
\end{subequations}
The temporal scalings of $\Gamma(t)$ and $P(t)$ are given by the ultra-relativistic jump conditions across the shock, such that taking $g(1) = f(1) = h(1) = 1$ implies $d \log \Gamma^2/ d\log t = -m$ and $d \log P/ d\log t = -m-k$. Substituting the definitions in Eq \eqref{eq:chi_def}--\eqref{eq:p_def_UR} into equation set \eqref{eq:hydro_ultra_rel}, we obtain the following self-similar equations
\begin{subequations}\label{eq:SelfSimEqs}
    \begin{equation}\label{eq:eq_selfsim_g}
        \frac{1}{g\chi}\frac{d \log g}{d \log \chi} = \frac{(7m+3k)-m~g\chi}{(m+1)(g^2\chi^2-8g\chi+4)}\equiv \eta_\g(g\chi)
    \end{equation}
    \begin{equation}\label{eq:eq_selfsim_f}
        \frac{1}{g\chi}\frac{d \log f}{d \log \chi} = \frac{4(2m+k)-(m+k)g\chi}{(m+1)(g^2\chi^2-8g\chi+4)}\equiv\eta_\f(g\chi)\,.
    \end{equation}
\end{subequations}
In the next section, we derive first and second type solutions to equation set \eqref{eq:SelfSimEqs}.

\section{Overview of First and second type self-similar solutions}\label{sec:FirstSecondTypeSol}
The results of this section reproduce \citealt{Sari2006}'s solutions to Eq \eqref{eq:SelfSimEqs}.
\subsection{First type solutions}\label{sec:FirstTypeSol}
Dimensional considerations imply that the explosion energy is a relevant parameter of the solution and the similarity index $m$ is found by requiring energy conservation. The energy of the shocked fluid has the following scaling
\begin{equation}
    E \sim \Gamma^2 \rho R \propto t^{1-k-m}\,.
\end{equation}
Therefore, for adiabatic blastwaves $m$ must satisfy
\begin{equation}\label{eq:m_type1}
    m_\I = 1-k\,.
\end{equation}
Solving equations \eqref{eq:SelfSimEqs} with $m = m_\I$, one obtains an analytic solution for the self-similar profiles :
\begin{subequations}
    \begin{equation}
        g = \chi^{-1}
    \end{equation}
    \begin{equation}
        f = \chi^{\frac{4k-7}{3(2-k)}}\,.
    \end{equation}
\end{subequations}
This solution is physical only if the energy in it converges. Integrating the energy density inside the blastwave gives
\begin{equation}
    E =  \int_0 ^{R_\sh} 4 p \gamma^2 dz \propto \int_1^\infty f g d\chi \propto \chi ^{\frac{4k-7}{3(2-k)}}\Big|_1^\infty\,.
\end{equation}
One immediately finds that first type solutions are restricted to $k<7/4$. The pressure gradient behind the shock is decreasing and becomes flat at $k=7/4$.
\subsection{Second type solutions}\label{sec:SecondTypeSol}
When the density profile becomes sufficiently steep, the shock accelerates rapidly and loses causal contact with the flow behind it. Loss of causal contact happens when the solution crosses a sonic line, defined by the locus of points on which $C_+$ characteristics, moving at the speed of sound with respect to the local fluid velocity, stay at a fixed self-similar coordinate (in the same way, a sonic line can be defined for $C_-$ characteristics that move in the negative $z$ direction with respect to the fluid). Energy conservation can no longer be imposed on the self-similar flow in order to solve for the similarity index $m$: the dynamics of the shock wave are independent of the part of the flow that contains most of the energy, which lies on the opposite side of the sonic point.
Instead, $m$ is found by eliminating the singularity from the equations.
Equation set \eqref{eq:SelfSimEqs} has two singular points $\chi_\pm$ that satisfy
\begin{equation}\label{eq:gchi_singular}
    g \chi_{\pm} = 4\pm2\sqrt{3}\,.
\end{equation}
The positive branch is associated with the sonic line of the negative $C_-$ characteristic, while the negative branch is the relevant one and defines the sonic line for a $C_+$ characteristic. Along $g\chi_+$, $C_+$ characteristics stay at the same coordinate $\chi$. Barring the formation of discontinuities, a well-behaved solution is required to pass smoothly through the sonic line. This constitutes the eigenvalue problem for $m$. Eliminating the singularity in equation set \eqref{eq:SelfSimEqs} requires having
\begin{equation}\label{eq:m_type2}
    m_\II = (3-2\sqrt{3})k\,.
\end{equation}
The solutions in this case are implicit in $g\chi$:
\begin{subequations}\label{eq:TypeIISol}
    \begin{alignat}{1}
        g &= \left[\frac{-g\chi+4+2\sqrt{3}-2\sqrt{3}k}{3+2\sqrt{3}-2\sqrt{3}k}\right]^{(2\sqrt{3}-3)k}\\
        f &= \left[\frac{-g\chi+4+2\sqrt{3}-2\sqrt{3}k}{3+2\sqrt{3}-2\sqrt{3}k}\right]^{(2\sqrt{3}-4)k}\,.
    \end{alignat}
\end{subequations}
In contrast to first type solutions, here the energy is allowed to diverge since most of it is carried by the non self-similar flow, which is out of causal contact with the shock. One should keep in mind, however, that the eigenvalue problem is defined only if the solution crosses the sonic line. In the absence of a singularity the solution must be consistent with energy convergence and conservation. In \S \ref{sec:AsymptoticPLSolution} we will show that second type solutions can be found only for $k>2$. One can immediately see that a problem arises when $7/4<k<2$, where the flow does not obey either of the known types of self-similarity. If the energy in this regime diverges, one must first solve for the non self-similar flow as it dictates the dynamics of the shock.

In order to study energy convergence, we obtain in \S \ref{sec:AsymptoticPLSolution} the asymptotic properties of the flow in the far downstream of the shock.

\section{Power-law asymptotic at $\chi \gg 1$}\label{sec:AsymptoticPLSolution}
In this section we investigate the asymptotic behaviour of the blastwave at $\chi \gg 1$ ($x \gg 0$) and find that $p$ and $q$ attain power-law profiles in $x$ and $t$.
To acquire some intuition, let us consider a fluid element that is passed by the shock at time $t_0$. Assuming that its Lorentz factor changes like $\gamma_\fe^2 \propto t^{-m_\fe}$ at $t \rightarrow \infty$, the distance between the shock and the fluid parcel at time $t$ is
\begin{equation}
    \Delta z = \int_{t_0}^t (\beta_\sh-\beta_\fe)dt \simeq 
    \left[\frac{t}{2(m_\fe+1)\gamma_\fe^2}-\frac{t_0}{2(m+1)\Gamma_0^2}\right]\,,
\end{equation}
where $\Gamma_0$ is the Lorentz factor of the shock at $t=t_0$. Since fluid elements do not overtake the shock ($m_\fe>-1$), the first term dominates and we find that $\gamma_\fe^2 \simeq \frac{t}{2(m_\fe+1)x}$ at $t\gg t_0$, where we approximate $\Delta z \simeq x$. Thus, away from the immediate vicinity of the shock $\gamma$ is a function only of $t$ and $x$ and is independent of $\Gamma_0$ (see also \citealt{Faran2021}). Using equations \eqref{eq:chi_def}, \eqref{eq:gamma_def} and the approximation for $\gamma_\fe^2$ we can expect to find that $g\chi \rightarrow constant$ in the limit $\chi \gg 1$. The asymptotic value of $g\chi$ can be found from Eq \eqref{eq:eq_selfsim_g} by rewriting it in the following form:
\begin{equation}
    \frac{d \log g\chi}{d \log \chi} = 1+\eta_\g(g\chi) \cdot g\chi = \frac{(g\chi - g\chi_1)(g\chi-g\chi_2)}{(g\chi-g\chi_+)(g\chi-g\chi_-)}\,,
\end{equation}
where 
\begin{equation}\label{eq:gchi_const}
    g\chi_{1,2} = \frac{8-3k+m \pm \sqrt{\mathcal{D}}}{2}
\end{equation}
are the two possible asymptotic values of $g\chi$, and we define the discriminant 
\begin{equation}
    \mathcal{D} \equiv-16(m+1)+(8-3k+m)^2\,.
\end{equation}
Generally, there are two asymptotic values for $g\chi$ and the solution settles on the one it encounters first, which depends on the boundary conditions at $\chi = 1$. A single asymptotic value exists if one of $g\chi_{1,2}$ coincides with $g\chi_\pm$, or if $\mathcal{D}=0$. The former case corresponds to second-type solutions, where one of the asymptotic values is equal to $g\chi_+$.
Substituting Eq \eqref{eq:m_type1} or Eq \eqref{eq:m_type2} for $m$ into Eq \eqref{eq:gchi_const} returns the following asymptotic solutions:
\begin{equation}\label{eq:gChi_as_m12}
    g\chi_\as = \begin{cases}\frac{9-4k \pm |7-4k|}{2}&\,,\mbox{$m=m_\I$}\\   4-\sqrt{3}k-\sqrt{3}|2-k|&\,,\mbox{$m=m_\II$}\,.
    \end{cases}
\end{equation}
The limiting behaviour of first type solutions can be alternatively derived in the following way: let us write the temporal change in energy through a surface of constant $\chi$:
\begin{equation}
        F(\chi) = \left(w \gamma^2-p\right)\frac{dz}{dt}\Big|_\chi - w \gamma^2 \beta\Big|_z  = p(1-g\chi) \,,
\end{equation}
where $w = 4p$ is the enthalpy.
Integrating the energy conservation equation over a constant self-similar volume implies $d F(\chi)/d\chi = dE(\chi)/dt$, where $E(\chi)$ is the energy contained in the interval $d\chi$. This equality translates to
\begin{equation}
        (1-g\chi)\eta_\f =  -\frac{1-m-k}{m+1}\,.
\end{equation}
Since Eq \eqref{eq:m_type1} enforces $(1-g\chi)\eta_\f = 0$, the two possible values for $g\chi$ are either $g\chi = 1$, giving zero energy flux, or $g\chi = 4(2-k)$, satisfying constant pressure ($\eta _\f = 0$). Eq \eqref{eq:gChi_as_m12} reduces to these two values when taking $k\geq7/4$. Imposing zero flux at the shock forces the solution to settle on $g\chi = 1$. An analogous interpretation for second type solutions is less straightforward.

Second type solutions exist only when the solution crosses the sonic line, i.e., $g\chi_\as<g\chi_+$. Equations \eqref{eq:gchi_singular} and \eqref{eq:gChi_as_m12} show that this happens only for $k>2$, such that $g\chi_\as = 4+2\sqrt{3}(1-k)$.

The asymptotic hydrodynamical profiles are found by substituting Eq \eqref{eq:gchi_const} into equation set \eqref{eq:SelfSimEqs}. We obtain power-law profiles with exponents that depend on $m$ and $k$:
\begin{equation}\label{eq:PowerLawSol}
    q = Q \frac{t}{x}\,, ~ p\propto t^{-\alpha}x^\beta\,,
\end{equation}
where
\begin{equation}\label{eq:alpha_beta}
    \alpha = \frac{1}{3}(4-Q^{-1})\,, ~ \beta = \frac{4}{3}(Q-1)\,,
\end{equation}
and
\begin{equation}\label{eq:Q_m}
    Q \equiv \frac{1}{8(m+1)}\left[m+8-3k + \sqrt{\mathcal{D}}\right]\,.
\end{equation}
It is clear from Eq \eqref{eq:PowerLawSol} - \eqref{eq:alpha_beta} that the energy diverges at $x\rightarrow\infty$ if $\beta\geq0$ ($Q>1$).
We have already shown that the energy in first type solutions converges for $k<7/4$. Second type solutions have $Q>1$ and therefore the energy in them diverges asymptotically. Nevertheless, this is not a problem since most of the energy is carried by the non self-similar part of the flow (not described by Eq \ref{eq:TypeIISol}), which is causally disconnected from the shock. In reality, the solution beyond the sonic point deviates from that of the self-similar flow such that the energy in the blastwave converges. One can therefore remain oblivious to the properties of the inner flow when solving for the dynamics of the shock. 

Although $m$ and $Q$ are unknown when $7/4<k<2$, upper and lower limits can be set on $m$. The asymptotic solution necessarily satisfies $g\chi_\as = \max\{g\chi_1,g\chi_2\}$ and also $g\chi_\as>g\chi_+$. These two conditions give $m\geq(3-2\sqrt{3})k$. Another restriction comes from the requirement that $\mathcal{D}\geq0$, which is satisfied for $m\leq-4\sqrt{3(k-1)}+3k$ or $m\geq4\sqrt{3(k-1)}+3k$. Only the first inequality is consistent with a continuous solution for $m$ as a function of $k$. Therefore, $m$ must lie in within $(3-2\sqrt{3})k\leq m\leq -4\sqrt{3(k-1)}+3k$. These limits imply that inside the gap $Q>1$ and the energy diverges asymptotically at $x\rightarrow\infty$. One therefore faces a new challenge: the energy diverges, while a sonic point is not part of the solution. This necessarily means that the self-similar dynamics are dictated by the non self-similar part of the flow, which contains most of the energy. In order to study the dynamics of the shock one must first solve for the non self-similar flow.

A clue to how one should address the problem comes from the fact that when the pressure increases behind the shock, the asymptotic solution satisfies $p(0)=0$ and $q(0)\rightarrow\infty$. This kind of behaviour is expected in the leading edge of a relativistically hot gas that expands into vacuum. We therefore expect the non self-similar flow to follow expansion into vacuum, and look for exact solutions that coincide with the power law asymptotic of Eq \eqref{eq:PowerLawSol} at $x\rightarrow0$. We impose initial conditions in which the energy in the flow converges, and then check what kind of shocks can be driven by these flows. The similarity index $m$ will then be found from the condition for the coexistence of the shock with the fluid that expands into vacuum behind it. The associated self-similar solutions will naturally be consistent with energy conservation and convergence.  

\section{Ultra-relativistic expansion into vacuum}\label{sec:ExpIntoVacuum}
In this section we obtain exact solutions describing the expansion of an ultra-relativistic gas into vacuum.
Our goal is to find solutions in which the flow approaches the powerlaw asymptotic (Eq \ref{eq:PowerLawSol}) expected to appear at $x \rightarrow 0$, while deviating from it at some large $x$ in order to comply with energy convergence.

In the first stage, we reduce the two fluid equations \eqref{eq:hydro_ultra_rel} to a single PDE by performing a hodograph transformation. We proceed by finding solutions that approach the powerlaw asymptotic at $x\rightarrow0$, and in the final stage construct a solution that satisfies specific boundary conditions imposed by initial profiles with finite energy.

\subsection{Hodograph transformation}
The fluid equations \eqref{eq:hydro_ultra_rel} can be further simplified via a hodograph transformation.
We change independent variables from $\{t,x\}$ to $\{u,v\}$:
\begin{subequations}
    \begin{equation}\label{eq:u_qp_def}
        u = -\frac{\sqrt{3}}{12}\ln\left[q  p^{\sqrt{3}/2}\right]
    \end{equation}
        \begin{equation}\label{eq:v_qp_def}
        v = \frac{\sqrt{3}}{12}\ln\left[q  p^{-\sqrt{3}/2}\right]\,,
    \end{equation}
\end{subequations}
where $u$ and $v$ are the Riemann invariants of the plane parallel fluid equations in the ultra-relativistic limit (see e.g., \citealt{JohnsonMckee1971}), conserved along $C_+$ and $C_-$ characteristics, respectively.
The numerical prefactor $\sqrt{3}/12$ is introduced to simplify the equations obtained below. The pressure and Lorentz factor can be represented in terms of the new variables by
\begin{eqnarray}
    p &=& \e^{-4(u+v)}\,,\label{eq:p_uv}\\
    q &=& \e^{-2\sqrt{3}(u-v)}\label{eq:pq_uv}\,.\label{eq:q_uv}
\end{eqnarray}
The parameters $u$ and $v$ are defined in the entire space $-\infty<u,v<\infty$ and correspond to $0<p,q<\infty$.
We make an additional transformation to a new dependent variable $\psi(u,v)$:
\begin{equation}\label{eq:PsiPhi}
    \psi = \left(\frac{q}{p}\right)^{-1/2} \Phi = \e^{-u(2-\sqrt{3})-v(2+\sqrt{3})}\Phi\,,
\end{equation}
where $\Phi(u,v)$ is defined by its partial derivatives
\begin{equation}\label{eq:t_chi}
    t = p \cdot \Phi_p = -\frac{1}{8}\left(\partial_u+\partial_v\right)\Phi\,,
\end{equation}
and
\begin{equation}\label{eq:x_chi}
    x = \frac{p}{q} \Phi_p +\frac{3}{4}\Phi_q = -\frac{1}{16q}\left[\left(2+\sqrt{3}\right)\partial_u+\left(2-\sqrt{3}\right)\partial_v\right]\Phi \,.
\end{equation}
The above variable transformation brings equation set \eqref{eq:hydro_ultra_rel} into the form of a remarkably simple Klein-Gordon equation:
\begin{equation}\label{eq:KleinGordonEq}
    \psi_{uv} = \psi\,.
\end{equation}
Next, we find solutions to Eq \eqref{eq:KleinGordonEq} that approach the power law asymptotic of Eq \eqref{eq:PowerLawSol}.

\subsection{Solutions of the Klein-Gordon equation}
In this section, we find exact solutions to Eq \eqref{eq:KleinGordonEq} that will subsequently be used to describe expansion into vacuum.

It is easy to verify that 
\begin{equation}\label{eq:PsiPlSol}
    \psi_\pl(u\lambda,v/\lambda) = A \cdot\e^{\lambda u +v/\lambda}
\end{equation}
is a special solution of Eq \eqref{eq:KleinGordonEq} that returns the power law asymptotic of Eq \eqref{eq:PowerLawSol}, where the characteristic parameter $\lambda$ determines the powerlaw indices $\alpha$ and $\beta$ through its relation to $Q$:
\begin{equation}\label{eq:LambdaQ}
    \lambda = \frac{-1+2Q(2+\sqrt{3})}{2+\sqrt{3}-2Q}\,,
\end{equation}
and $A$ is a constant that only affects the pressure amplitude.

Eq \eqref{eq:KleinGordonEq} is also solved by the function $\psi = F(u\lambda,v/\lambda)$, defined as
\begin{equation}\label{eq:PsiExactSolFuv}
    F(u,v) = \sum\limits_{n = 1}^{\infty} \frac{u^n}{n!} \sum\limits_{k = 0}^{n-1}\frac{v^k}{k!}\,.
\end{equation}
In the limit $u\gg1$ and $v\gg1$, $F(u \lambda,v/\lambda)$ simplifies to
\begin{equation}\label{eq:F_uv}
    F(u\lambda,v/\lambda) \simeq \begin{cases}
        \e^{u\lambda+v/\lambda} + \frac{I_0(2\sqrt{uv})}{\sqrt{v/(u \lambda^2)}-1}  &\,,\mbox{$u \gg v/\lambda^2$}\\
        \frac{I_0(2\sqrt{uv})}{\sqrt{v/(u \lambda^2)}-1}  &\,,\mbox{$u \ll v/\lambda^2$}\,,
    \end{cases}
\end{equation}
where $I_0(2\sqrt{uv}) \simeq \frac{\e^{2\sqrt{u v}}}{\sqrt{4\pi}(u v)^{1/4}}$ is the modified Bessel function $I_n(2\sqrt{uv})$ with $n=0$, in the limit $uv \gg 1$. Since $\e^{2\sqrt{u v}}\ll \e^{u\lambda+v/\lambda}$ when $u \gg v/\lambda^2$, $F(u \lambda, v/\lambda)$ coincides with the powerlaw asymptotic solution (Eq \ref{eq:PsiPlSol}) in that limit.

The problem defined by Eq \eqref{eq:KleinGordonEq} with boundary conditions on $C_\pm$ characteristics (corresponding to fixed values of $u$ and $v$) is a well posed problem, which can be translated to an initial condition problem in $p$ and $q$. The solution to Eq \eqref{eq:KleinGordonEq} is then a superposition of $\psi_\pl$ and $F(u \lambda,v/\lambda)$ (and any other additional solution of Eq \ref{eq:KleinGordonEq}) that satisfies the boundary conditions. 
We obtain the solution that develops from specific initial profiles in the next subsection.

\subsection{Solution for specific initial conditions}
The results of the previous subsections will now be used to achieve our stated goal: finding a solution that contains finite energy and asymptotes to the powerlaw profiles of Eq \eqref{eq:PowerLawSol} at $x\rightarrow 0$. For that, we impose initial conditions in which the energy converges: in the region $0<x<x_0$, the pressure is an increasing power-law of $x$, represented by $\psi_\pl(u \lambda,v/\lambda)$ with $\lambda>2+\sqrt{3}$. At $x>x_0$ the pressure is a decreasing power-law of $x$, also represented by $\psi_\pl(u \mu,v/\mu)$ with a different index $\mu<2+\sqrt{3}$. This profile ensures energy convergence at $x \rightarrow \infty$. 
The two power-law solutions intersect at the point $x_0$, at which the Riemann invariants are chosen (without loss of generality) to be $u = v = 0$, such that $q(x_0)=p(x_0)=1$. The initial profiles at time $t_0$ are therefore
\begin{equation}\label{eq:pq_InitialConditions}
    p(t_0) = \begin{cases}
        (x/x_0)^{\beta(\lambda)}&\,,\mbox{$x<x_0$}\\
        (x/x_0)^{\beta(\mu)} &\,,\mbox{$x>x_0$}
    \end{cases}\,, ~~~ q(t_0) = x_0/x\,,
\end{equation}
where $\beta(\lambda)>0$ and $\beta(\mu)<0$.
In terms of $\psi(u,v)$, the initial conditions are given by Eq \eqref{eq:PsiPlSol}:
\begin{equation}\label{eq:PsiInitialConditions}
    \psi_0 = \begin{cases}
        A_\lambda \cdot \e^{u\lambda + v/\lambda} + \Delta\psi_\lambda&\,,\mbox{$u<0, v>0$}\\
        A_\mu \cdot \e^{u\mu + v/\mu} + \Delta\psi_\mu &\,,\mbox{$u>0, v<0$}
    \end{cases}\,,
\end{equation}
where the terms $\Delta \psi_\lambda$ and $\Delta \psi_\mu$ are introduced to account for the offsets $x \rightarrow \Delta x$ and $t \rightarrow \Delta t$, that together with $A_\lambda$ and $A_\mu$ are required for constructing the initial profiles of $p$ and $q$ (see Appendix \ref{A:PsiTranslations}).

\begin{figure}
    \centering
    \includegraphics[width = 0.98\columnwidth]{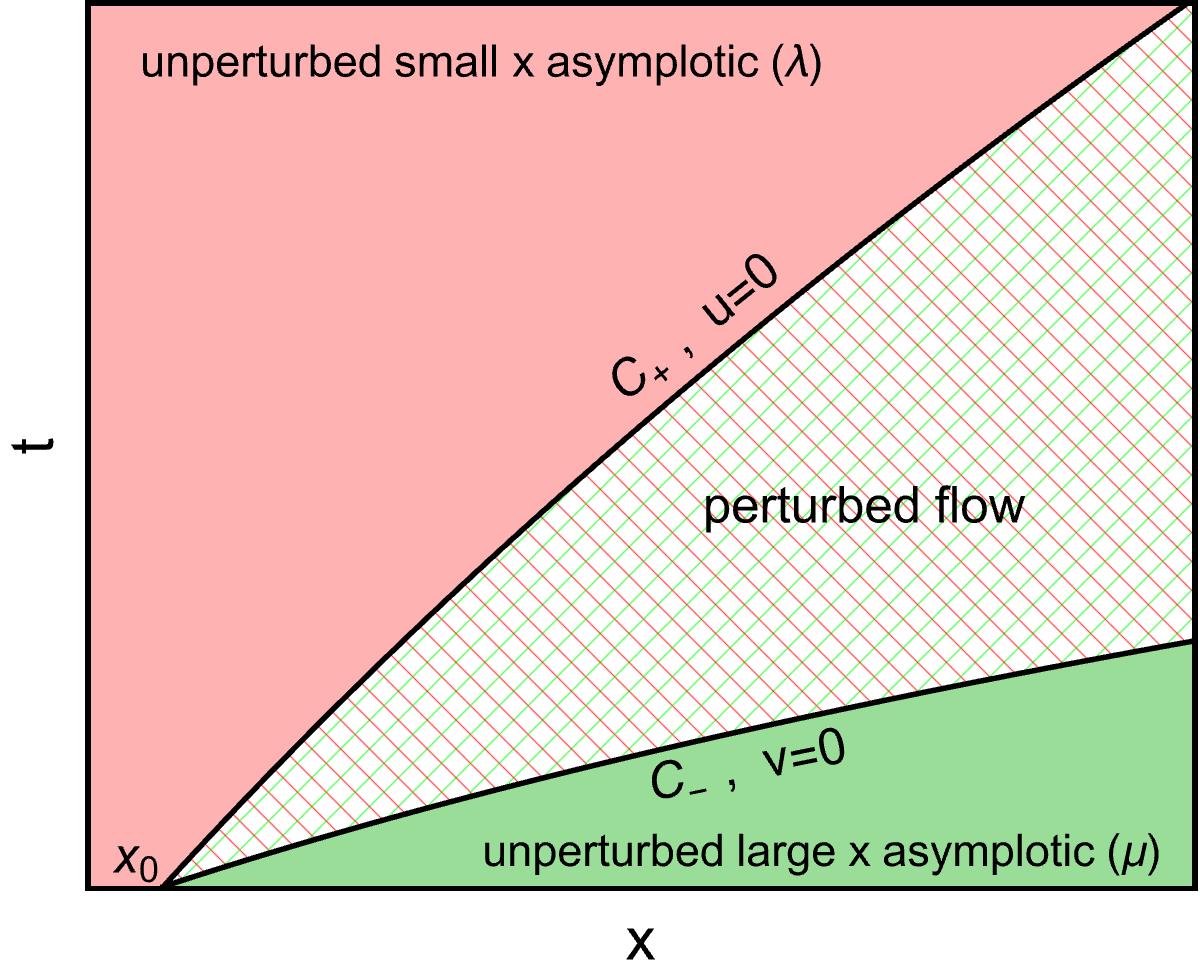}
    \caption{A schematic characteristic plot illustrating the different regions that develop from the initial conditions of Eq \eqref{eq:pq_InitialConditions} at times $t>t_0$. The flow is perturbed between the limiting $C_+$ and $C_-$ characteristics that emerge from $x_0$, while outside this region it retains its exact initial profiles.}
    \label{fig:char_perturbations}
\end{figure}

Perturbations to the initial profiles occur upon the arrival of the limiting $C_\pm$ characteristics that emerged from $x_0$ at the initial time $t_0$. The $C_+$ characteristic along which $u=0$ propagates within the small x power-law asymptotic, while the $C_-$ characteristic carrying the value $v=0$ propagates within the larger-x asymptotic.
As illustrated in Fig \ref{fig:char_perturbations}, the perturbed flow lies in the region between the two characteristics, while outside of it the fluid retains its original profiles. The boundary conditions of Eq \eqref{eq:KleinGordonEq} are therefore defined on the limiting $C_\pm$ characteristics that emerge from $x_0$. Eq \eqref{eq:KleinGordonEq} needs to be solved only in the first quadrant, where $u, v>0$; everywhere else the flow is unperturbed and the solution is given by Eq \eqref{eq:PsiInitialConditions}.

The problem thus admits the boundary conditions $\psi(0,v) = A_\lambda \cdot \e^{v/\lambda} + \Delta \psi_\lambda(0,v)$ and $\psi(u,0) = A_\mu \cdot \e^{u\mu} + \Delta \psi_\mu (u,0)$. 
Although one can write down a solution that satisfies the exact boundary conditions, doing so requires adding multiple terms to $\psi$ that are only important for small values of $u$ and $v$. Recalling that we are interested in the asymptotic evolution of the flow at $t \gg t_0$ where $u, v \gg 1$, the boundary conditions can be simplified by taking the limit of $u, v \rightarrow \infty$, thus neglecting the contribution of the subleading terms $\Delta \psi _\lambda$ and $\Delta \psi _\mu$:
\begin{equation}\label{eq:BoundaryConditions}
    \psi(u = 0,v)\simeq A_\lambda \cdot \e^{v/\lambda}\,, ~ \psi(u,v=0) \simeq A_\mu \cdot \e^{u\mu}\,.
\end{equation}
We therefore solve the problem defined by Eq \eqref{eq:KleinGordonEq} and \eqref{eq:BoundaryConditions}, which provide an accurate description of the problem in the limit $t \rightarrow \infty$. We find the following solution for $u,v>0$:
\begin{equation}\label{eq:PsiSolExact}
\begin{split}
        \psi &= A_\lambda \left[\e^{u\lambda + v/\lambda} - F(u\lambda,v/\lambda) \right]+A_\mu F(u\mu,v/\mu) \\
        &\simeq\begin{cases}
            A_\lambda \e^{u\lambda + v/\lambda}&\,,\mbox{$u\ll v/\lambda^2$}\\
            A_\mu \e^{u\mu + v/\mu}&\,,\mbox{$u\gg v/\mu^2$}
        \end{cases}\,.
    \end{split}
\end{equation}
Interestingly, the limits taken in the second equality of Eq \eqref{eq:PsiSolExact} show that the same powerlaw asymptotics that were imposed on the initial profiles form independently in parts of the perturbed flow. The boundaries of the powerlaw asymptotics thus do not lie along the axes but rather along the lines $u = v/\lambda^2$ and $u = v/\mu^2$, whose respective characteristic widths are $\Delta u_\lambda = 2\sqrt{v/\lambda^3}$ and $\Delta u_\mu = 2\sqrt{v/\mu^3}$. Despite the fact that perturbations from the large $x$ asymptotic are carried into $0<u<v/\lambda^2$ along $C_+$ characteristics, the flow there remains \emph{effectively} undisturbed. The same is true for the large x asymptotic, where the region $0<v<u\mu^2$ remains effectively unchanged to perturbations arriving from the small x asymptotic along $C_-$ characteristics.


Using Eq \eqref{eq:F_uv} in the limit $u \gg v/\lambda^2$, it can be wshown that between the two powerlaw asymptotics, the solution is
\begin{multline}\label{eq:PsiIntermediate_Region}
    \psi(v/\mu^2 -\Delta u_\mu< u < v/\lambda^2+\Delta u_\lambda) \simeq \\ \frac{\e^{2\sqrt{u v}}}{\sqrt{4 \pi}(u v)^{1/4}}\left(\frac{A_\lambda}{\sqrt{v/(u\lambda^2)}-1}-\frac{A_\mu}{\sqrt{v/(u\mu^2)}-1}\right)\,.
\end{multline}
We point out that the flow in this regime has negligible dependence on $\lambda$ and $\mu$, as they do not appear in the exponent. In what follows, we will show that this part of the flow constitutes a piston that contains most of the energy in the solution.

\begin{figure}
    \centering
    \includegraphics[width = 1\columnwidth]{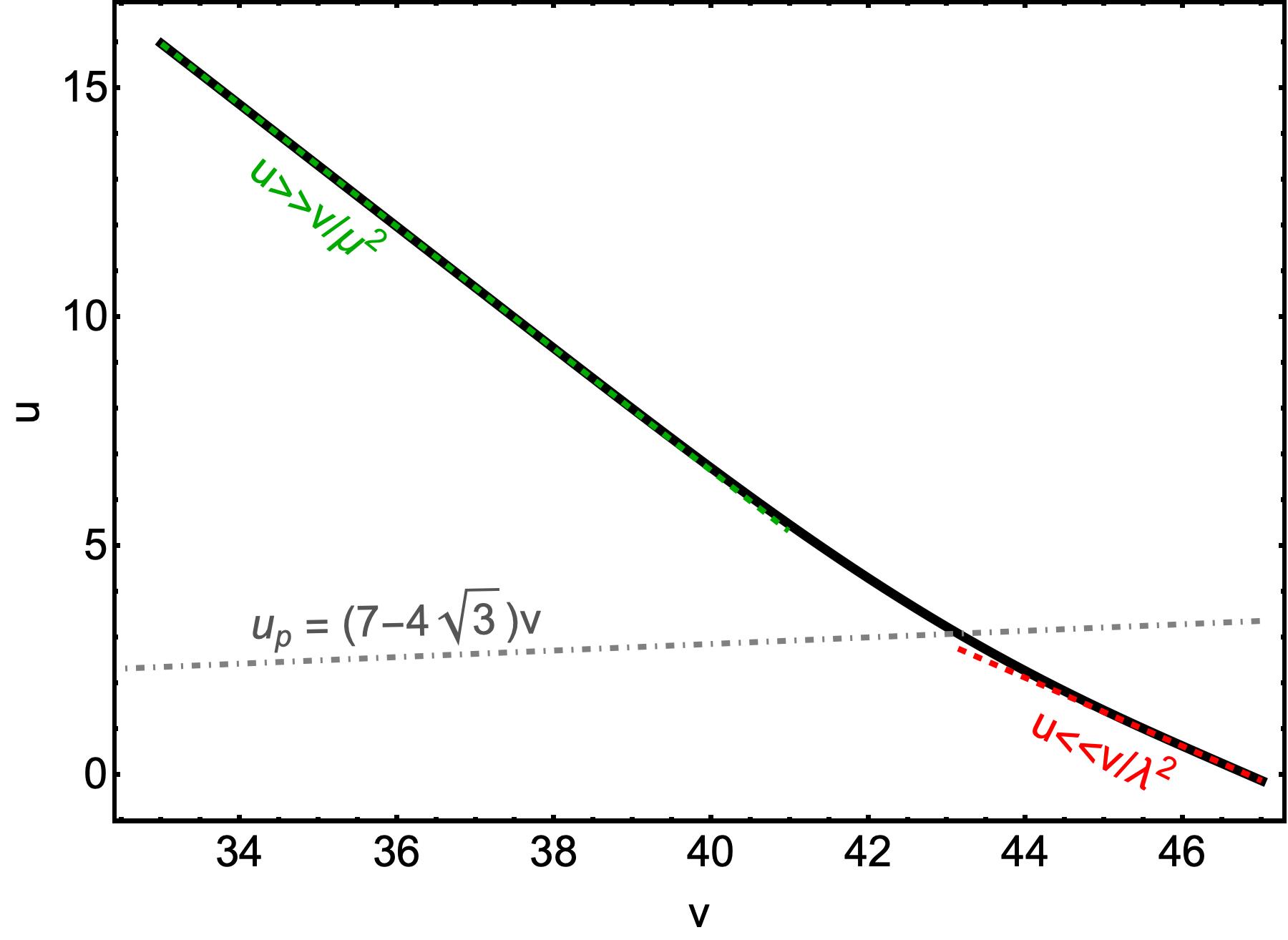}
    \caption{Exact solution obtained from Eq \eqref{eq:PsiSolExact} (black solid line) at time $t = 10^{80}$ using $\lambda = 5$ and $\mu = 2.8$. The asymptotic power-law solutions of the small and large x asymptotic are the red dashed and green dashed lines, respectively. The line tracing the pressure maximum is also plotted (gray dot-dashed line).}
    \label{fig:sol_t1e80_uv}
\end{figure}

We use equations \eqref{eq:t_chi} and \eqref{eq:PsiSolExact} with $x_0 = 1$, $\lambda = 5$ and $\mu = 2.8$ to solve for $u(v)$ at time $t = 10^{80}$. Figure \ref{fig:sol_t1e80_uv} shows that in the limits $u \ll v/\lambda^2$ and $u \gg v/\mu^2$ the exact solution coincides with those of the small and large x asymptotics, respectively. Corresponding profiles of $p(x)$ and $q(x)$ are shown in Fig \ref{fig:sol_t1e80_Pq} using equations \eqref{eq:p_uv}, \eqref{eq:q_uv}, \eqref{eq:t_chi} and $\eqref{eq:x_chi}$.

\subsection{Ultra-relativistic piston}
We are interested in the properties of the flow in the intermediate region given by Eq \eqref{eq:PsiIntermediate_Region}.
Using equations \eqref{eq:PsiPhi}, \eqref{eq:t_chi} and \eqref{eq:PsiIntermediate_Region} in the limits $u,v \rightarrow \infty$, one obtains a solution for $u(v,t)$ that is independent of $\lambda$ and $\mu$:
\begin{equation}\label{eq:uv_IntermediateSol}
    u = (2+\sqrt{3})\left[\ln(t)^{1/2}-\sqrt{v(2+\sqrt{3})}\right]^2\,.
\end{equation}
Since the solution lies between the powerlaw asymptotics of increasing and decreasing pressure, a maximum must be described by Eq \eqref{eq:uv_IntermediateSol}.
It can be seen from Eq \eqref{eq:p_uv} that a maximum in pressure is described by the requirement $du/dv = -1$. Taking the full derivative of Eq \eqref{eq:uv_IntermediateSol} with respect to $v$ at constant $t$, we have
\begin{equation}
    \frac{du}{dv}\left[(2-\sqrt{3})+\sqrt{\frac{v}{u}}\right]+\sqrt{\frac{u}{v}}+2+\sqrt{3} = 0\,.
\end{equation}
Setting $du/dv = -1$ and solving the quadratic equation (keeping in mind that $u$ and $v$ are real and therefore only one root is relevant), we find that the line that traces the pressure maximum on the $\{u,v\}$ plane is
\begin{equation}\label{eq:up_v_0}
    u_\p = (7-4\sqrt{3})v\,.
\end{equation}
The temporal scaling of $v$ at the peak is found by substituting Eq \eqref{eq:up_v_0} into Eq \eqref{eq:uv_IntermediateSol}:
\begin{equation}\label{eq:vp_t_0}
    v_\p = \frac{\ln(t)}{16(2-\sqrt{3})}\,.
\end{equation}
Finally, we find the scaling of $q$ and $p$ at the peak by substituting equations \eqref{eq:up_v_0}-\eqref{eq:vp_t_0} into equations \eqref{eq:p_uv}-\eqref{eq:q_uv}:
\begin{equation}\label{eq:pp_t0}
    p_\p = \e^{-16\left(2-\sqrt{3}\right)v} = t^{-1}
\end{equation}
\begin{equation}\label{eq:qp_t0}
    q_\p = \e^{12(2-\sqrt{3})v} = t^{3/4}\,.
\end{equation}
The flow around the peak contains most of the energy and constitutes a hot, accelerating piston.
It is interesting to note that at the lower boundary of the solution gap ($k=7/4$) , the boundary of the powerlaw asymptotic satisfies $v/\lambda^2 = u_\p(v)$, which implies that powerlaw asymptotic extends all the way to the peak of the pressure. This result is reassured by the fact that the temporal scaling relations of Eq \eqref{eq:pp_t0}--\eqref{eq:qp_t0} are identical to those of the shock's for $k=7/4$.
When $k>7/4$, the profiles transitions from the power-law asymptotic to the intermediate solution before the peak.

\begin{figure}
    \centering
    \includegraphics[width = 1.0\columnwidth]{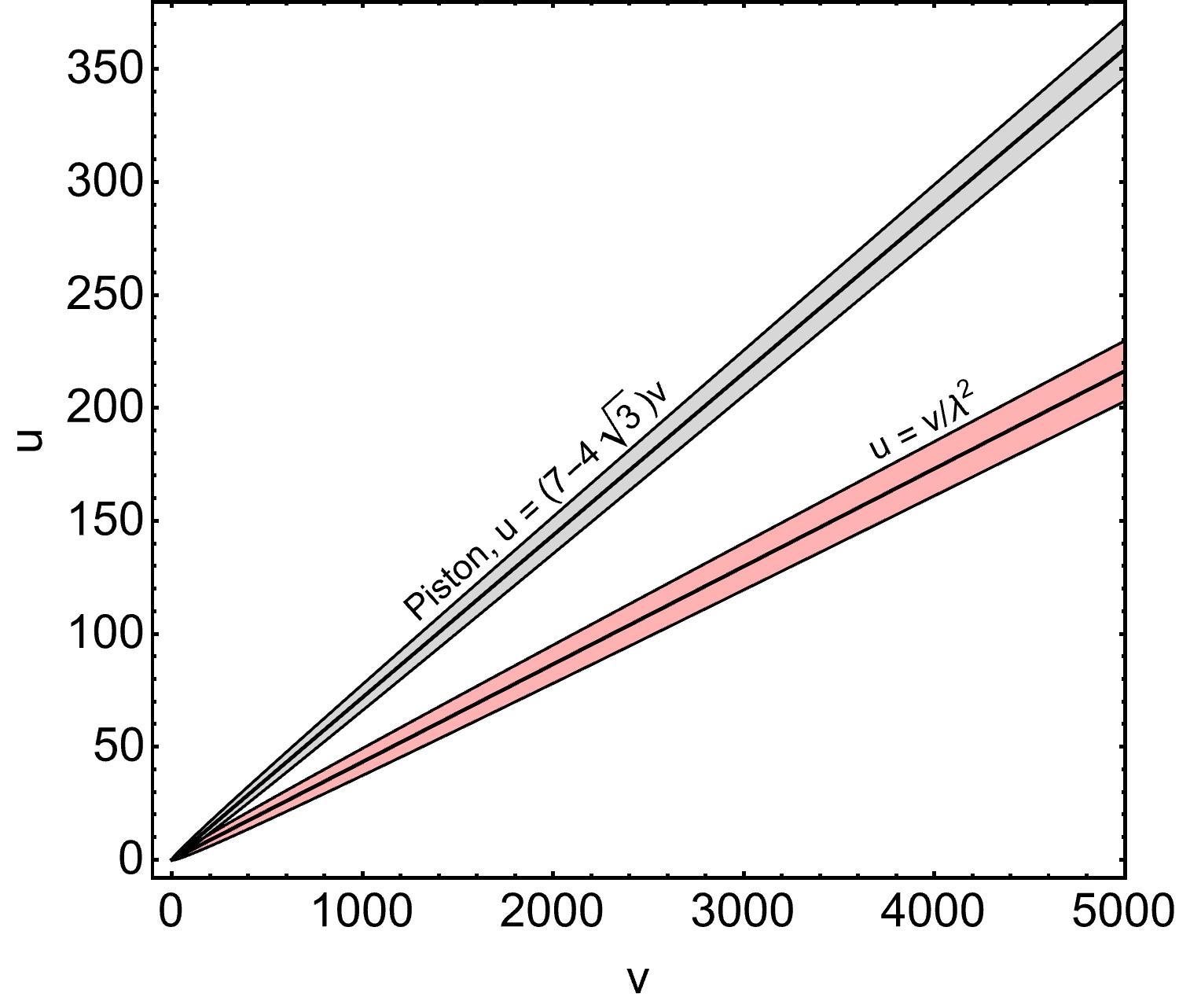}
    \caption{The two components of expansion into vacuum. The line $u = v/\lambda^2$ is the boundary between the power-law asymptotic and the rear solution and the respective width over which the solution changes is shaded in red. The peak of the pressure is traced by the line $u = (7-4\sqrt{3})v$, and the shaded gray region designates the width of the peak.}
    \label{fig:uv_plot}
\end{figure}

The width of the pressure maximum on the $\{u,v\}$ plane can be estimated by $\Delta v_\p = \left(\frac{p}{d^2p/dv^2}\right)^{1/2}\Big|_{v_\p}$, yielding a narrowing relative width of $\Delta v_\p/v_\p= \sqrt{2}(2-\sqrt{3})\ln(t)^{-1/2}$ and $\Delta u_\p/u_\p = \sqrt{2}(2+\sqrt{3})\ln(t)^{-1/2}$. The two characteristic lines $u = v/\lambda^2$ and $u_{\p}(v)$ are shown in Figure \ref{fig:uv_plot}.

\begin{figure}
    \centering
    \includegraphics[width = 0.98\columnwidth]{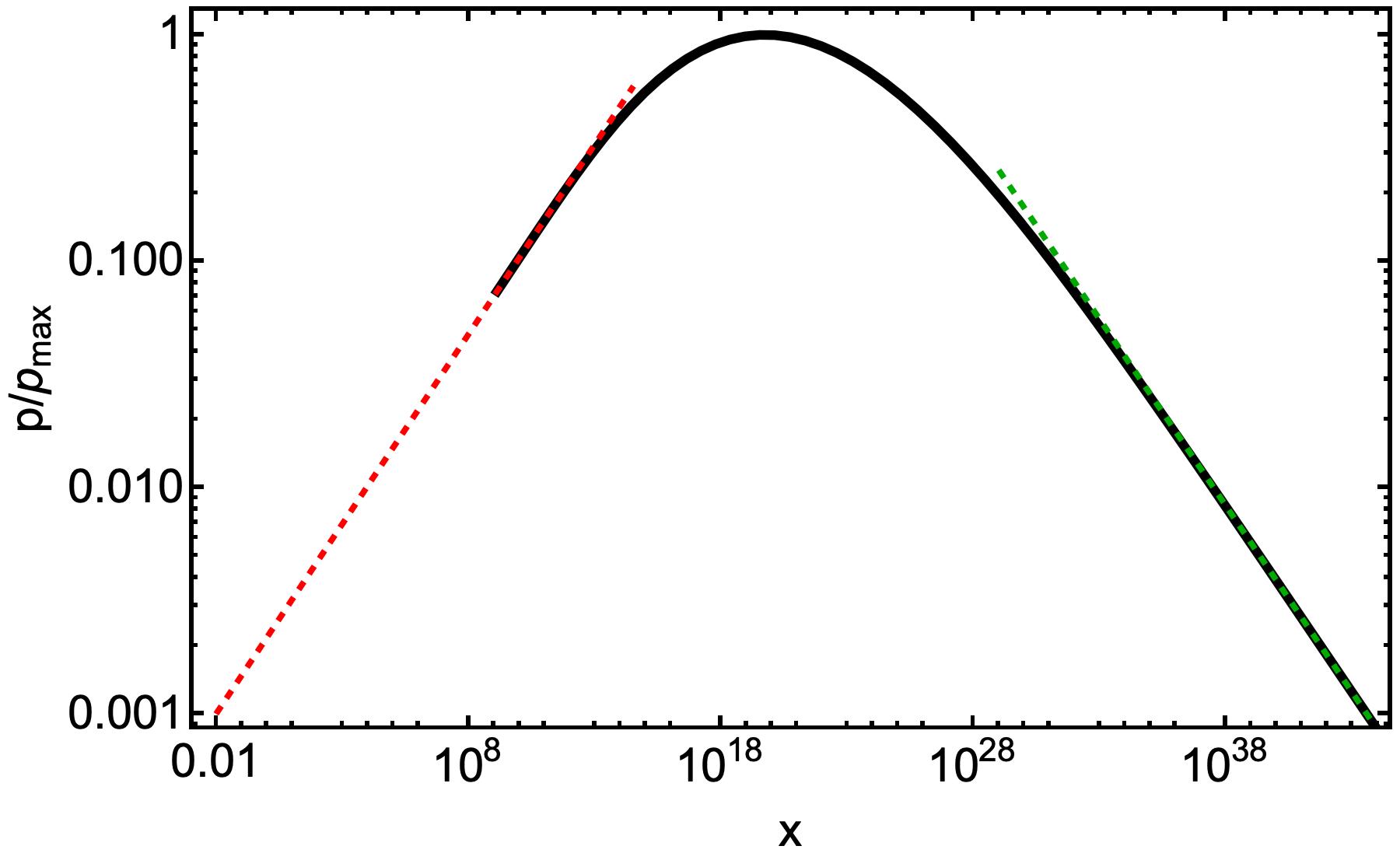}
        \includegraphics[width = 0.975\columnwidth]{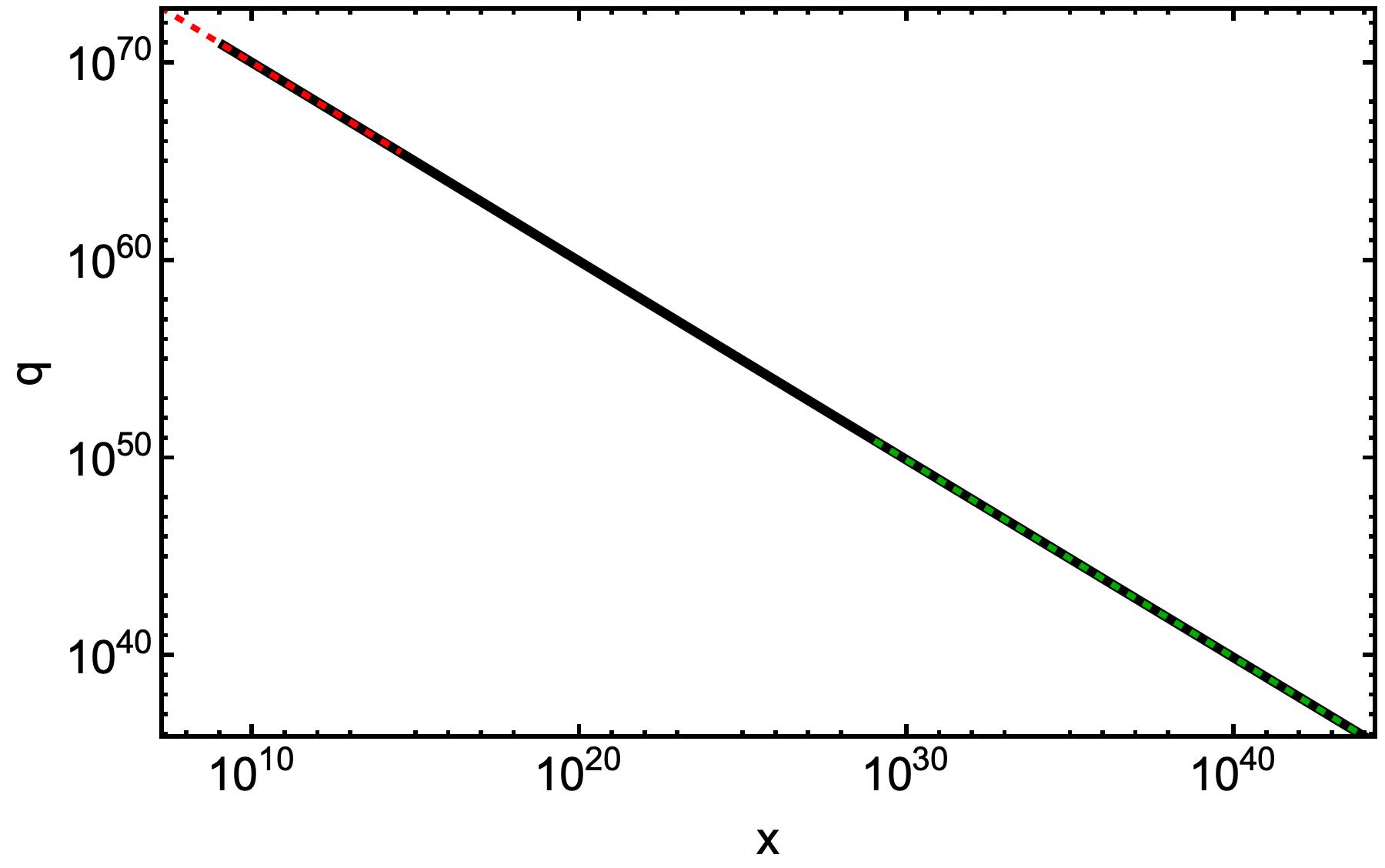}
    \caption{Exact profiles of $p$ and $q$ obtained from Eq \eqref{eq:PsiSolExact} (black solid lines) and asymptotic power-law solution of the small x (red dashed line) and large x (green dashed line) asymptotics at $t = 10^{80}$. The solution at small x terminates at $u=0$ and is extended into the fourth quadrant, where $u<0$ and $v>0$ using the unperturbed profiles set by the initial conditions.}
    \label{fig:sol_t1e80_Pq}
\end{figure}

The Lorentz factor of the flow can be written in a form similar to $q = Q t/x$ of the power-law solution, while the equivalent of $Q$ is now a function of $u$ and $v$:
\begin{multline}\label{eq:q_tx_IntermediateSol}
    q = \frac{t}{x}\left[1-\frac{\sqrt{3}}{16}\frac{\left(\partial_u-\partial_v\right)\Phi}{t}\right] \\= \frac{t}{x}\left[\frac{u(2-\sqrt{3})+v(2+\sqrt{3})+2\sqrt{u v}}{2(u+v+4\sqrt{uv})}\right]\,.
\end{multline}
We confirm that substituting $u = v/\lambda^2$ returns $q = Q(\lambda)t/x$, as expected. Eq \eqref{eq:q_tx_IntermediateSol} also suggests that the profile of $q$ around the piston should not show considerable deviation from those in the powerlaw asymptotics, as observed in Fig \ref{fig:sol_t1e80_Pq}; while $t/x$ depends exponentially on $u$ and $v$, its prefactor has only linear dependence.

The position of the pressure maximum is found by substituting Eq \eqref{eq:up_v_0} into Eq \eqref{eq:q_tx_IntermediateSol}, thus obtaining
\begin{equation}
    x_\p(t) = \frac{t}{q} = \frac{t_0}{q_{\p,\i}}\left(\frac{t}{t_\i}\right)^{1/4}\,,
\end{equation}
with $t_\i$ being some reference time at which $q_\p(t_\i) = q_{\p,\i}$.
While the relative widths of the pressure maximum in terms of $\Delta v_\p/v_\p$ and $\Delta u_\p/u_\p$ decrease with time, the spatial relative width of the peak increases logarithmically:
\begin{equation}
    \frac{\Delta x_\p}{x_\p} = \frac{\sqrt{3/2}}{2}\ln(t)^{1/2}\,.
\end{equation}
We confirm that the energy in the solution is conserved by recalling that most of it is carried by the piston, whose energy can be estimated by $E \sim q_\p p_\p \Delta x_\q \propto t^{-1}t^{3/4} t^{1/4} = const$, where the typical scale height of the energy density, $\Delta x_\q \sim x_\p \propto t^{1/4}$, is dictated by the profile of $q$.

Further Intuition about the nature of the piston can be gained by a heuristic description of a hot, relativistic blob that undergoes free expansion, while exchanging internal energy with bulk kinetic energy. Let us associate $\Gamma_\b$ with the bulk motion of the blob, and $\gamma_\th$ with the random motion of its particles (thermal energy). Since $E \sim \Gamma_\b(\gamma_\th m c^2) = constant$ due to energy conservation, we have $\gamma_\th \propto\Gamma_\b^{-1}$. Adiabatic expansion implies $\gamma_\th m c^2 \propto (\Gamma_\b \Delta )^{-1/3}$, where $\Delta \propto t/\Gamma_\b ^2$ is the volume of the expanding blob in the lab frame. This gives $\Gamma_\b^2 \propto t^{1/2}$. The scaling of the pressure can be found from energy conservation: $E \sim \Gamma_\b ^2 p (t /\Gamma_\b^2) = p\cdot t = const \longrightarrow p\propto t^{-1}$. The evolution of the pressure is inferred solely from energy conservation and therefore agrees with our solution for the piston.
However, this analysis returns a different scaling for $q$, which immediately implies that the expansion of the piston is not adiabatic. We conclude that the piston's entropy is not conserved, which suggests that it is not occupied by the same material throughout its evolution. 


\begin{figure}
    \centering
    \includegraphics[width = 1\columnwidth]{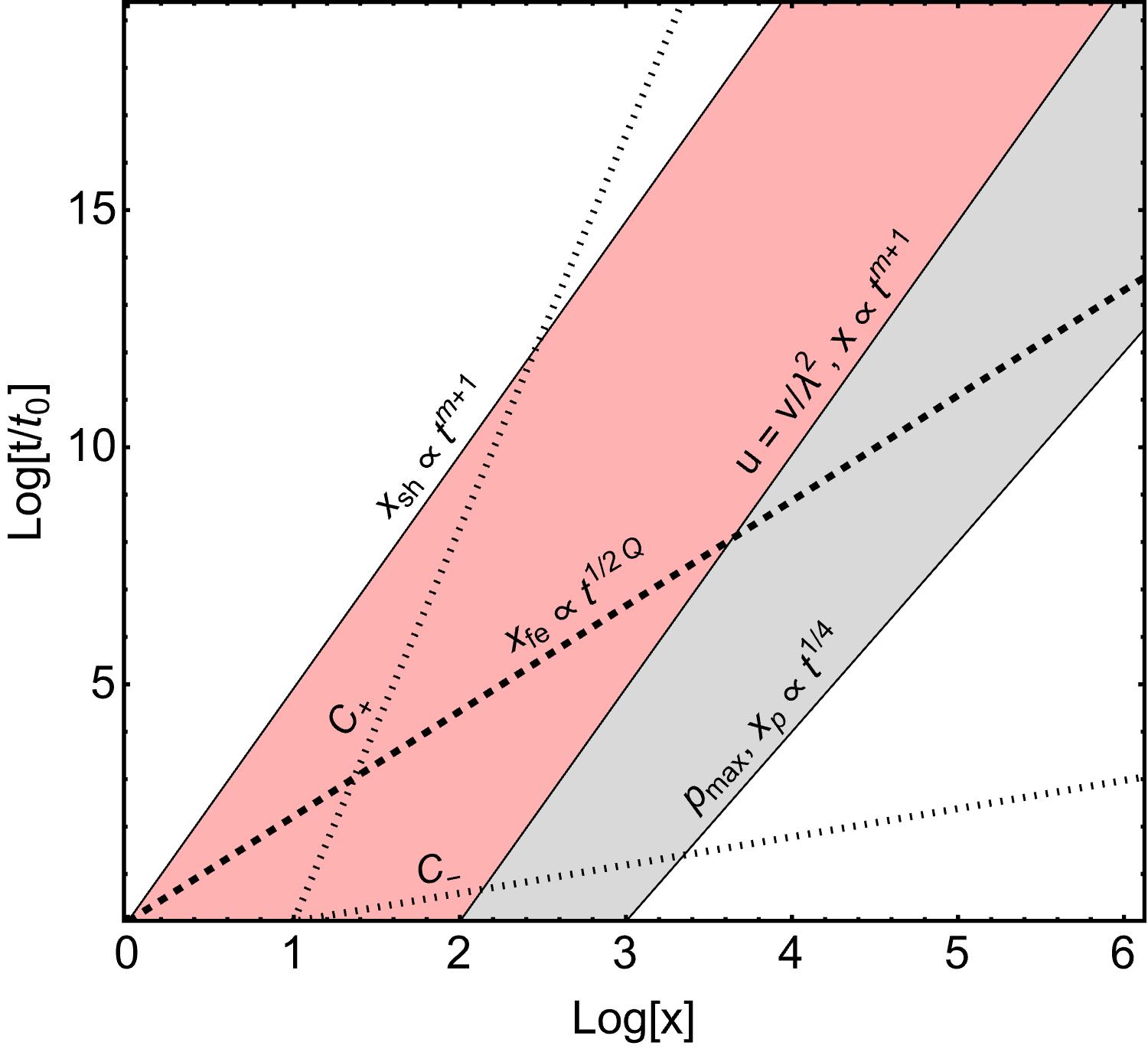}
    \caption{An illustration of the blastwave's dynamics. The power-law asymptotic region that develops behind the shock is shown in red, and the region between the powerlaw boundary and the peak of the pressure is in gray. We plot the trajectories of $C_\pm$ characteristics emerging from the power-law asymptotic behind the shock that demonstrate that the flow is in causal contact. A fluid element, whose position is designated by $x_\fe$ starts at the shock and joins the piston at later times.
    The dashed lines assume scalings of the asymptotic power-law solution. }
    \label{fig:xt_plot}
\end{figure}

\section{Consistency with a shock}\label{sec:Shock}
In the previous section we obtained an exact solution for relativistic gas expanding into vacuum, whose leading front approaches the powerlaw asymptotic expected to form in the far downstream of a strong shock. Our objective is to smoothly connect the rear part of the self-similar blastwave solution with the leading front of expansion into vacuum. In order for both solutions to coexist with one another, one must find the appropriate values of the temporal index $m$ for which the powerlaw asymptotic that forms by the piston at $u,v>0$ is not destroyed by the shock.

The above condition is more easily interpreted by considering the trajectories of the shock and the boundary of the powerlaw asymptotic on the $\{u,v\}$ plane. A shock can form self-consistently from expansion into vacuum if $du/dv|_\sh<1/\lambda^2$, so that it never crosses the line $u = v/\lambda^2$. Using the ultra-relativistic shock jump conditions, this requirement translates to the following inequality:
\begin{equation}\label{eq:ShockInequality}
    \frac{du}{dv}\Big|_\sh = \frac{2m+\sqrt{3}(m+k)}{-2m+\sqrt{3}(m+k)} \leq \frac{1}{\lambda^2}\,.
\end{equation}
At the same time, as $Q$ (and therefore $\lambda$) depends on the discriminant $\mathcal{D}$ (see Eq \ref{eq:Q_m}), an additional constraint comes from the requirement $\mathcal{D}
\geq0$. Assuming $\lambda$ is finite, we find that both conditions are satisfied only if $du/dv |_\sh = 1/\lambda^2$, which then gives the desired value of $m$:
\begin{equation}\label{eq:m_gap}
    m_\III = -4\sqrt{3(k-1)}+3k\,,
\end{equation}
thereby closing the gap between first and second type solutions. We confirm that both $m(k)$ and $m'(k)$ are continuous at $k=7/4$ and $k=2$, and note that $m_\III$ returns $\mathcal{D}=0$. A self-similar solution for the blastwave can now be found by solving the self-similar ODEs (Eq \ref{eq:SelfSimEqs}) with  $m=m_\III$.

One might expect that the interpretation of a shock driven by expansion into vacuum continues to be correct also in the regime of second type solutions, as the external density gradient becomes even steeper. Indeed, in solving the inequality in Eq \eqref{eq:ShockInequality} we have made the assumption that $\lambda$ is finite. However, in second type solutions all $C_+$ characteristics originate from the immediate vicinity of the sonic point, and therefore $u = constant$ in space. This corresponds to taking the limit $\lambda \rightarrow \infty$
for which Eq \eqref{eq:m_gap} returns $m = m_\II = (3-2\sqrt{3})k$. The analysis in this paper thus applies to all $k>7/4$.

A characteristic plot summarizing the main features of the flow is shown in Fig \ref{fig:xt_plot}. A fluid element that crossed the shock at an arbitrary time $t_0$ joins the asymptotic self-similar flow designated by the red region between the lines $x_\sh$ and $x_\pl$. As it is advected away from the shock, it leaves the self-similar flow after crossing $x_\pl$ and eventually joins the bulk of the fluid at the piston when reaching $x_\p$. A pair of $C_\pm$ characteristics are also plotted to demonstrate that the flow is in causal contact. The position of the shock and the boundary of the powerlaw asymptotic have the same scaling and therefore the distance between them increases with time.

\section{Direct Numerical Simulation}\label{sec:Simulations}

The problem introduced by equation set \eqref{eq:hydro_ultra_rel}
at $x>x_s(t)$, with boundary conditions
\be\label{bco}
q(t,x_s(t))=\frac{1}{4\dot{x}_s},~~~p(t,x_s(t))=\frac{t^{-k}}{3\dot{x}_s},
\ee
where $x_s(t)$ is the shock position, can be solved numerically as follows: introducing the Riemann invariants
\be
V=q^{-\frac{1}{2}}p^{-\frac{\sqrt{3}}{4}},~~~U=q^{-\frac{1}{2}}p^{\frac{\sqrt{3}}{4}}\,,
\ee
equations \eqref{eq:hydro_ultra_rel} can be written as
\be
\dot{V}+\left(1-\frac{\sqrt{3}}{2}\right)V U V'=0,
\ee
\be
\dot{U}+\left(1+\frac{\sqrt{3}}{2}\right)V U U'=0.
\ee
where overdot and prime represent partial derivative with respect to time and space, respectively. Making the following transformation of the independent variable
\begin{equation}
    x\rightarrow x-x_s(t)\,,
\end{equation}
we have for $x>0$:
\be\label{num1}
\dot{V}+\left[\left(1-\frac{\sqrt{3}}{2}\right) V U-Z\right]V'=0\,,
\ee
\be\label{num2}
\dot{U}+\left[\left(1+\frac{\sqrt{3}}{2}\right)V U-Z\right]U'=0\,,
\ee
with
\be
Z\equiv \frac{1}{4}V(t,0) U(t,0)\,.
\ee
Solving equations \eqref{num1}-\eqref{num2} requires only one boundary condition at $x=0$. This is because $1-\frac{\sqrt{3}}{2}<\frac{1}{4}<1+\frac{\sqrt{3}}{2}$
and $V$ propagates into the shock, while $U$ propagates out of the shock. It follows that only the $U$ boundary condition is needed. From the original boundary conditions (Eq \ref{bco}),
\be\label{num3}
U(t,0)=\left[\frac{4}{3}t^{-k}V(t,0)^{\frac{2}{\sqrt{3}}-1}\right]^{2\sqrt{3}-3}.
\ee
The numerical algorithm is now straightforward. We solve Eq \eqref{num1}-\eqref{num2} at $x>0$, propagating the Riemann invariants to the left or to the right, as dictated by the corresponding velocities. At the shock, at $x=0$, we compute $V(t,0)$ by propagating $V$ from the right, and then we compute $U(t,0)$ from the boundary condition (\ref{num3}). 

\begin{figure}
    \centering
    \includegraphics[width=1\linewidth]{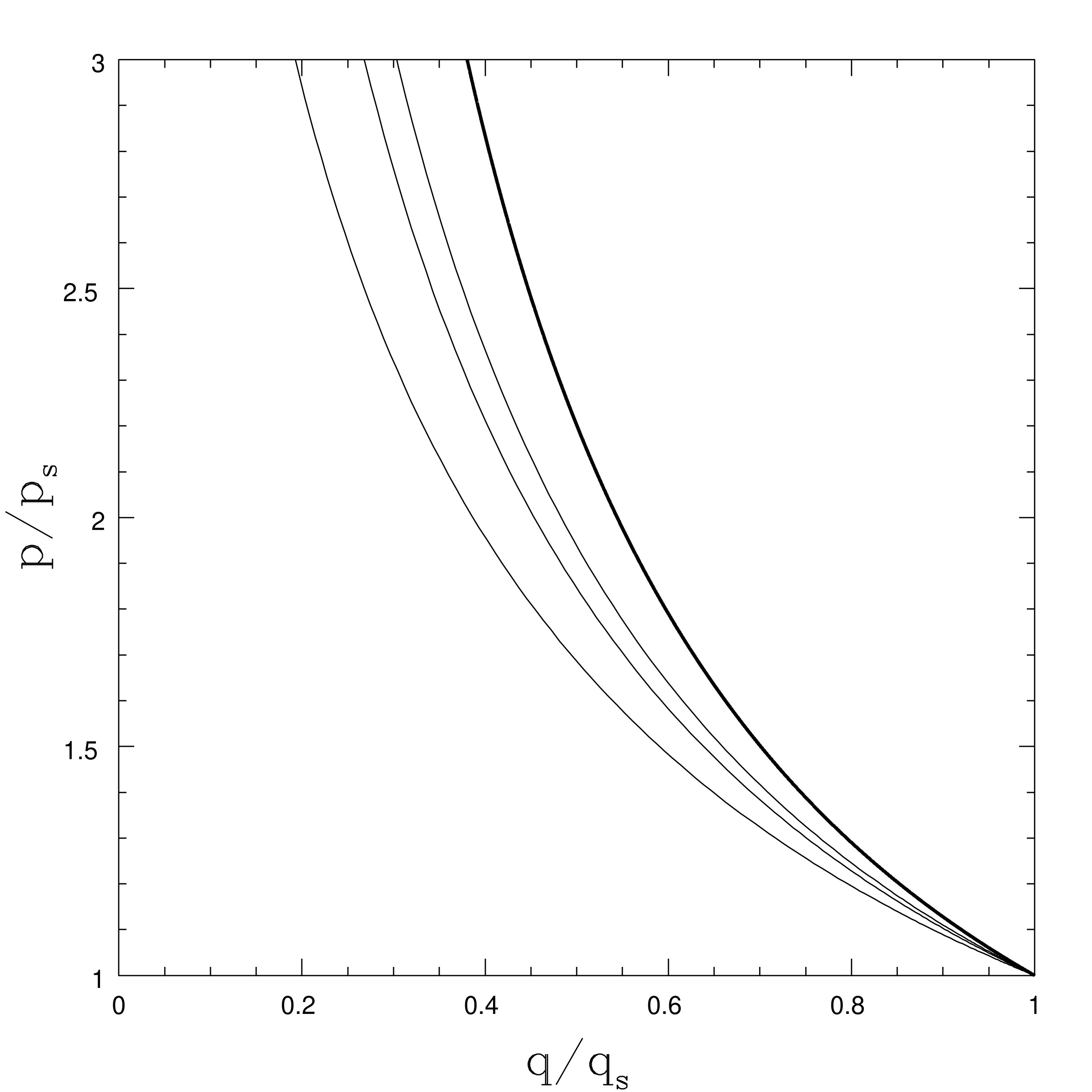}
    \caption{Numerical simulations of a shock propagating in a medium with $k = 1.98$. The profiles of $p$ vs. $q$ normalized to their values at the shock are plotted at times $t = 10^{15}, 10^{25}, 10^{35}$ and the theoretical curve is in black. The numerical profiles approach the theoretical curve, yet exact self-similarity is not reached by $t=10^{35}$ due to the slow convergence of the simulation.}
    \label{fig:numerics}
\end{figure}

We run the simulation for $k = 1.98$ with initial conditions at $t=1$ where a shock is placed at $x_\s = 1$ and the pressure behind it decreases as $p \propto x^{-0.0231}$. In Figure \ref{fig:numerics} we show the profiles of $p$ vs. $q$ at $t = 10^{15}, 10^{25}, 10^{35}$, together with the theoretical curve (thick line). The numerical profiles approach the theoretical curve as time increases. At $t = 10^{35}$ the numerical shock velocity satisfies $m = d \log \dot{x_\s}/ d\log t = -0.9138$, whereas the theoretical value given by Eq \eqref{eq:m_gap} is $m =-0.9186$. While our numerical results seem to agree with theory, the simulation converges very slowly and we are therefore not able to claim perfect agreement.

\section{Summary and Discussion}\label{sec:Summary}
We study the propagation of a plane-parallel, ultra-relativistic shock down a powerlaw density profile of the form $\rho \propto z^{-k}$. In the parameter space $7/4<k<2$ the blastwave cannot be described by any of the known types of self-similar solutions; global conservation laws do not apply in the self-similar domain, and at the same time no eigenvalue problem can be defined. Instead, the solution gap is populated by a new class of self-similar solutions. Following, we summarize our main results:
\begin{itemize}
\item
We find that within the solution gap the flow obeys self-similarity of the third type. The unique characteristic of this new class of solutions is that the similarity index $m$ is determined by the part of the flow that does not participate in the self-similar solution. This is a consequence of the fact that most of the energy in the blastwave resides in the non self-similar flow, which is in causal contact with the shock. Solving for the shock therefore requires a physical understanding of the solution in the entire space. In this work, we first obtained an exact solution for the non self-similar flow, described by relativistic expansion into vacuum, and then solved for the similarity index $m$ by requiring that the leading edge the expanding flow is not destroyed by the shock. We find that $m = -4\sqrt{3(k-1)}+3k$ for $7/4<k<2$.

\item 
The flow that expands into vacuum has two general components: (1) an accelerating piston that contains most of the energy and (2) a leading edge described by a powerlaw asymptotic in $t$ and $x$. The properties at the pressure maximum within the piston are characterized by universal exponents that are independent of $k$: $p_\p \propto t^{-1}$, $q_\p \propto t^{3/4}$ and $x_\p \propto t^{1/4}$. Despite containing most of the energy, the piston does not go through free expansion, so that its entropy is not conserved.

\item
Expansion into vacuum drives shock waves that obey second type self-similarity when $k>2$. Therefore, the analysis in this paper generally applies whenever $k>7/4$. Nonetheless, if $k>2$ solving for the non self-similar flow is not required to describe the shock and its near downstream; a loss of causal contact between the shock and the bulk of the blastwave means that the solution beyond the sonic point is unable to affect the shock, and is therefore not important for that purpose.

\item
Self similar solutions are important because they describe the asymptotic behaviour of a physical system for a large class of initial conditions. In addition, they also offer the technical benefit of having to solve a set of ODEs instead of a set of PDEs, which makes them a very useful tool. This simplifying feature was not used in this work, where an exact solution to the PDEs was obtained in order to derive the self-similar asymptotic thereafter. 
Nonetheless, the exact solution we find for expansion into vacuum does not probe the flow directly behind the shock, before it settles on the powerlaw asymptotic. In order to study the small scales behind the shock, one must indeed solve the self-similar equations \eqref{eq:SelfSimEqs} with the corresponding index $m(k)$.

\item
Numerical simulations seem to agree with our analytic results for the shock. However, due to the very slow numerical convergence we are not able to confirm perfect agreement with theory.

\item
We note that the original classification to first and second type solutions (\citealt{Barenblatt1976}, \citealt{Zeldovich1967}) only makes the distinction between problems that can be solved by dimensional considerations versus the need to solve an eigenvalue problem. The definition of second type self-similarity should be refined to include the requirement that the eigenvalue problem is imposed exclusively by the self-similar flow.

\item
Ultra relativistic blastwaves are often used in astrophysics in the context of Gamma-Ray Burst afterglows, in which spherical symmetry is imposed. According to \citealt{Sari2006}, in the spherically symmetric case both first and second type solutions can be found for $5-\sqrt{3/4}<k<17/4$, such that there is an overlap rather than a gap.
An overlap also appears in the plane parallel implosion problem (where $R \rightarrow 0$) between $7/4<k<1+\sqrt{3/4}$, which is relevant for describing relativistic shock breakout from a stellar surface. 
While numerical simulations can be used to break those degeneracies, one might be able to find exact solutions by generalizing this work to spherical symmetry or to plane-parallel implosions. We note that another solution gap exists in spherically symmetric implosions (see \citealt{Sari2006}), and is likely analogous to the plane parallel case.
\end{itemize}

\section*{Acknowledgements}
R.S. is partially supported by ISF, NSF/BSF and MOS grants.

\appendix
\section{Generalization of $\psi(u,v)$ to $x\rightarrow x+\Delta x$ and $t\rightarrow t+\Delta t$ } \label{A:PsiTranslations}
A temporal offset $t \rightarrow t+ \Delta t$ and a spatial translation $x \rightarrow x+ \Delta x$ can be introduced by adding the following term to $\psi(u,v)$:
\begin{equation}
\begin{split}
    \Delta \psi = & \frac{4}{\sqrt{3}} \left[u(2-\sqrt{3})-v(2+\sqrt{3})\right]\e ^{-u(2-\sqrt{3})-v(2+\sqrt{3})} \Delta t\\
    &+  \frac{4}{3}\e ^{-u(2+\sqrt{3})-v(2-\sqrt{3})}\Delta x\,.
\end{split}
\end{equation}
The initial conditions defined by Eq \eqref{eq:pq_InitialConditions} impose $\Delta t_\lambda = 0$ and $\Delta x_\lambda = \Delta x_\mu = 0$. The parameters $\Delta t_\mu,\, A_\lambda$ and $A_\mu$ in this case are
\begin{subequations}\label{eq:PsiInitialParameters}
    \begin{equation}
        \Delta t_\mu = \left(1-\frac{Q(\lambda)}{Q(\mu)}\right)t_0
    \end{equation}
    \begin{equation}
        A_\lambda  = -\frac{16 \lambda x_0}{(2+\sqrt{3})\lambda^2+2\lambda+2-\sqrt{3}}
    \end{equation}
    \begin{equation}
        A_\mu  = -\frac{16 \mu x_0}{(2+\sqrt{3})\mu^2+2\mu+2-\sqrt{3}}
    \end{equation}
\end{subequations}
where $t_0 = -A_\lambda(4+\lambda+1/\lambda)/8$ and $Q(\lambda), Q(\mu)$ are given by Eq \eqref{eq:LambdaQ}.
The exact boundary conditions are then $\psi(0,v) = A_\lambda \cdot \e^{v/\lambda}$ and $\psi(u,0) = A_\mu \cdot \e^{u\mu}+ \frac{4 \Delta t_\mu}{\sqrt{3}} u(2-\sqrt{3})\e ^{-u(2-\sqrt{3})}$, where the subleading term that appears due to $\Delta t_\mu$ vanishes in the limit $u\rightarrow\infty$.
\bibliography{ThirdTypeURSol.bib}

\end{document}